# Revealing the low-temperature fast relaxation peak in a model metallic glass

B. Wang[a,b,$], L. J. Wang[c,d,$], B. S. Shang[c], X. Q. Gao[e], Y. Yang[f,*], H. Y. Bai[a,g,h], M. X. Pan[a,g,h,*], W. H. Wang[a,g,h], P. F. Guan[c,*]

[a]Institute of Physics, Chinese Academy of Sciences, Beijing 100190, China

[b]School of Physical Science and Technology, Northwestern Polytechnical University, Xi'an 710072, China

[c]Beijing Computational Science Research Center, Beijing 100193, China

[d]School of Physics and Materials Science, Anhui University, Hefei 230601, China.

[e]Northwest Institute for Nonferrous Metal Research, Xi'an 710016, China

[f]Department of Mechanical Engineering, City University of Hong Kong, Tat Chee Avenue, Kowloon Tong, Kowloon, Hong Kong, China

[g]School of Physical Sciences, University of Chinese Academy of Sciences, China

[h]Songshan Lake Materials Laboratory, Dongguan, Guangdong 523808, China

Corresponding Author

*Y. Y email: yonyang@cityu.edu.hk

*M. X. P email: panmx@aphy.iphy.ac.cn

*P. F. G email: pguan@csrc.ac.cn

**ABSTRACT**

By systematically investigating the relaxation behavior of a model metallic glass based on the extensive molecular dynamics (MD) simulations combined with the dynamic mechanical spectroscopy method, a pronounced ultra-low temperature peak on the loss modulus spectrum was discovered for the first time in MD simulations. It was found that the relation peak occurs at a much lower temperature than the typical temperature for the conventional β relation peak as reported in the literature. According to the atomic displacement analysis, we unravel that the reversible atomic motions, rather than the thermal vibrations or local structural rearrangements, mainly contribute to this relaxation peak. We further identify the atomic level mechanism of this fast relaxation process by characterizing the local geometrical anisotropy. Furthermore, by tracing the dynamic behaviors of these "reversible" atoms, we demonstrate the intrinsic hierarchy of the local relaxation modes, which are triggered by atomic vibrations and gradually developed to the reversible and irreversible atomic movements. Our findings shed light on a general picture of the relaxation processes in metallic glasses.

**Keywords:** metallic glasses; relaxation dynamics; molecular dynamics simulation; fast relaxation; reversible motions.

# 1. Introduction

In condensed matter physics, the relaxation dynamics of glasses is one of important and challenging but yet unsolved issues [1-5]. For structurally complicated glasses, such as polymeric and molecular glasses, a complex relaxation spectrum exists, displaying salient peaks in typical relaxation experiments, for example, the α peak, the β peak, even the γ peak and so on [6,7]. In contrast, due to the seemingly random packing of atoms and the lack of rotational degrees of freedom [8-10], metallic glasses (MGs) were once deemed to have a very simple relaxation behavior, *i.e.*, there could be only a single α peak that involves large-scale irreversible structural rearrangements on the relaxation spectrum. However, numerous studies found that MGs could also exhibit a secondary β peak [11]. This relaxation mode has relation with shear transformation zones (STZs) [12], atomic diffusion [13] and dynamical heterogeneity [14], and has a pivotal influence on the tensile plasticity [15] and fragility [16] of MGs. Hence, the discovery of β relaxation in MGs has recently motivated lots of related research in the glass area [17,18]. Today, the β relaxation is often perceived to be governed by local structural rearrangements [19], acting as a precursor process to the α relaxation.

Recently, compelling experimental evidence has indicated the existence of an additional relaxation peak [20-22], which emerges at a lower temperature than that of traditional *β* relaxation in a number of MGs. The finding of this low temperature relaxation (LTR) peak is quite intriguing, which implies that there may be a trigger process to the *β* relaxation. If so, unraveling the origin of the LTR peak becomes critical, which may help resolve the long perplexing issues in the MG field, such as the onset of plasticity and boson peak [21,23-25]. Unfortunately, there lacks direct evidence to justify any atomic mechanisms proposed so far in the literature [20-22].

In this work, we aim to uncover the atomistic origin of the LTR peak through the study of a model $Cu_{50}Zr_{50}$ glass. Our results clearly demonstrate that this peak is due to the reversible activated motion of some atoms within their first-neighbor shells (FNSs). Compared to other atoms, these atoms participating in the reversible activated motion exhibit strong local geometric anisotropy. More importantly, we also show direct evidences that these reversible atomic motions indeed trigger a locally cooperative atomic sub-diffusive process, which governs the normal $\beta$ relaxation later observed at a higher temperature.

## 2. Simulation Methods

*2.1 Model system*

The model metallic glass system $Cu_{50}Zr_{50}$ (in atomic percent) we studied contained 4000 atoms interacting with an embedded atom method potential [26], and the molecular dynamics (MD) simulations were performed using the open source LAMMPS [27]. We apply periodic boundary conditions throughout the simulations during sample preparation, and the Nosé-Hoover thermostat was used to control temperature [28]. We first equilibrated the high temperature liquid at 1900 K for 10 ns, then quench the liquid to 50 K at a rate of $10^{10}$ K/sec in the NPT ensemble, where NPT ensemble in this article means constant number of particles $N$, pressure $P = 0$, and temperature $T$.

*2.2 Molecular dynamics simulation of dynamic mechanical spectroscopy*

The LTR behaviors of this glassy sample were investigated using extensive MD simulations. We load a strain $\varepsilon_{xy}(t) = \varepsilon_A \sin(2\pi t/t_p)$ along *xy* direction of samples, with $t_p$ presenting the period of strain and $\varepsilon_A$ meaning the maximum value of strain. Then, the phase shifts between loading strain and corresponding stress were collected for further analysis [29,30]. In the simulations, we set $t_p$=1000 ps and $\varepsilon_A$ =1% (which is in the linear elastic region) for dynamic mechanical spectroscopy (MD-DMS) of the main body. For each MD-DMS, 16 full cycles were loaded and data were collected during the latter 10 cycles to eliminate the artificial unstable local atomic packing and to ensure the system has reached the steady state under the sinusoidal loading. The function $\sigma_{xy}(t) = \sigma_0 + \sigma_A \sin(2\pi t/t_p + \delta)$ were used to fit the time dependent stresses. Then, the maximum resultant stress $(\sigma_A)$ and the phase shift between the strain and the stress $(\delta)$ are obtained. To get better statistics, we combined the MD-DMS and isoconfigurational ensemble methods together. Therefore, 50 independent MD-DMS loadings were carried out, starting from the same initial configuration but with momenta randomly assigned according to the Maxwell-Boltzmann distribution at the interested temperatures [31,32]. A typical MD-DMS measurement were shown in Fig. 1, where T=500 K and $t_p = 1000$ ps. And Fig. 2(a) shows the fitted parameters, $\delta$ and $\sigma_A$, at different temperatures. Finally, $E' = \frac{\sigma_A}{\varepsilon_A}\cos(\delta)$ and $E'' = \frac{\sigma_A}{\varepsilon_A}\sin(\delta)$ were used to calculate the storage modulus and loss modulus, respectively.

*2.3 Atomic pinning method*

The atomic pinning method [33-35], which only restricts selected atoms to move affinely with the simulation box during cycling, was employed to understand the contribution of the specific atoms. For a given configuration, we fix the positions of specific particles, arriving at a "frozen template". The specific (pinned) particles are only allowed to move affinely with the simulation box, while the remaining (unpinned) particles move as normal. Therefore, we can measure the MD-DMS spectroscopy through pinning some specific atoms interested and then repeating the above MD-DMS procedure to evaluate the effect of some specific atoms on $E''$. In this article, the pinned atoms are the reversible atoms which contribute to the low temperature peak or the same number of random vibrational atoms which vibrate around their initial residences.

## 3. Results and discussion

### 3.1 Discover of the fast relaxation dynamics

On the basis of the systematic MD-DMS simulations, the phase shift ($\delta$) between the applied cyclic strain and the stress response and the resultant stress amplitude ($\sigma_A$) are presented as functions of temperature in Fig. 2(a). Figure 2(b) shows the calculated loss modulus, $E'' = \frac{\sigma_A}{\varepsilon_A}\sin(\delta)$, as a function of temperature. The $E''$ curve exhibits a peak located at ~ 900 K, which corresponds to the α-relaxation. In other words, that signals the sample transform from the glassy state to supercooled liquid state [30].

Unexpectedly, a pronounced non-monotonic variation of $E''(\mathrm{T})$ can be observed in the temperature range from 100 K to 500 K, which has never been seen in the previous computational studies [19,30,36]. In contrast to the conventional $\beta$ relaxation emerging usually in the temperature range of (0.75-0.9) $T_g$ (the glass transition temperature) reported by the experimental and other numerical works [13,17,19,37], this peak is located at a much lower temperature (~0.3 $T_g$). To explore the underlying atomic response, three characteristic temperatures, including 100 K and 500 K that bound the non-monotonic $T$ dependence and 300 K, the peak temperature of the unusual LTR, were selected for the following atomic dynamics analysis. As suggested in the previous works [30,38], we calculated the mean square atomic jump distance $u(\Delta t) = ||\vec{r}(t_0 + \Delta t) - \vec{r}(t_0)||$ to characterize the dynamics of each atom, where $\Delta t = T_p = 1000$ ps and $T_p$ is the sinusoidal period of the applied strain. The probability density functions $p(u)$ of the resultant $u$ are shown in Fig. 2(c). Previous studies [30] reported that the "faster atoms" with $u > u_c$ (i.e. ~ 1.4 Å) might be responsible for LTR and the fraction of these atoms, $f(u > u_c)$, was supposed to be correlated with $E''$. However, as shown in Fig. 2(d), no correlations can be observed between the $f(u > u_c)$ and the $E''$ in the temperature range of 100 K-500 K. It implies that the pronounced $E''$ peak below 500 K is not governed by the "faster atoms" and the related relaxation process is not the one as discussed in previous studies [14,30]. Thus, the peak, denoted as $\beta'$ peak in Fig. 2(b), can be taken as the lower temperature fast relaxation, or fast $\beta$ relaxation peak. Despite the fact that this kind of LTR peak was observed in a variety of MGs experimentally [20,22], its atomistic origin is still mysterious.

*3.2 Correlation of reversible motions and the fast relaxation dynamics*

Apparently, tracking the jump distance $u(\Delta t = T_p)$ alone may not be sufficient to uncover the atomistic mechanism underlying the peak. To extract more comprehensive information, the trajectory of each atom was calculated, which is represented by a non-affine displacement, $u'(t) = \left|\left|\vec{r}(t) - \vec{r}(6T_p) - \int_{6T_p}^{t} \dot{\varepsilon} y(t)\hat{x} dt\right|\right|$ with $\dot{\varepsilon} = 2\pi\varepsilon_A/T_p \cos(2\pi t/T_p)$, over the time range, $t \in [6T_p, 16T_p]$. Compared with Fig. 2(c), we

can see there indeed exists some slight differences between the two calculation methods. As shown in Fig. 3(a), irrespective of the similarity that the number of atoms with large displacement increases as temperature increases, the notable feature can be seen that $p(u')$ has four distinct peaks. A closer examination of data in Fig. 3 uncovers that the first and second peaks are within first neighbor shell indicating that motions below 300 K behave as pronounced "cage effect"; the positions of latter two peaks have a direct correlation with the structure of MGs, matching the first and second peaks of $g_{Cu\text{-}Cu}(r)$ as indicated in Fig. 3(b), which is consistent with that copper atoms move faster in $Cu_{50}Zr_{50}$ MGs [5,8]. Considering $g(r)$ describes the static structure of system while $p(u')$ shows a dynamical behavior, the match between them implies the relations between dynamic and structural properties, and one atom can jump to the position previously taken by its nearest or secondary neighbors [19].

By systematically analyzing the features of $u'$, the atomic motions can be classified into three types. As shown in Fig. 4(a) and Fig. S1, Type I atoms simply vibrate smoothly around the initial locations with $u' < 1$ Å and therefore can be defined as the "vibrational atoms (VAs)". By comparison, the trajectories of type II and III atoms are intermittent, being composed of a succession of atomic motions from simple vibrations around well-defined locations to rapid reversible movements and ultimately to irreversible "jumps". Thus, the type II atoms are defined herein as the "reversible atoms (RAs)" with the maximum value of $u'_{max} \in (1\ \text{Å}, 2\ \text{Å})$, and the type III atoms as the "faster atoms (FAs)" with $u'_{max} > 2$ Å, which contribute to a diffusion-like process. These distinct features can be characterized through the distribution of $u'(t)$ over the investigated time range with the probability density functions $p(u')$ shown in Fig. 3(a). The differences between $p(u')$ and $p(u)$ send a strong message that the analysis solely based on the jump distance $u(\Delta t = T_p)$ ignores the important information about the reversible motion in the time interval $T_p$. According to the feature of $u'$, we extracted the fraction of each type of atoms during cyclic loadings and the results are shown in Fig. 4(b). It was found that most of the atoms belong to type I at a very low temperature (< 300 K) and the related fraction $f_{VA}$ decreases as temperature increases. However, the fraction of FAs, $f_{FA}$, almost equals 0 for T < 300 K but increases with further temperature rising. This finding indicates that the "FAs" cannot be the origin of the $\beta'$ peak. However, it was surprising to observe that the fraction of RAs, $f_{RA}$, exhibits a non-monotonic variation with temperature, echoing very well with the profile of the $\beta'$ peak in Fig. 2(b).

By performing 50 independent MD-DMS simulations based on the isoconfigurational ensemble, we found that the average number of RAs, $N_{RA}$, as shown in the inset of Fig. 5(a), presents a non-monotonic variation with temperature, which is consistent with the temperature profile of $E''$. Figure 5(a) shows an excellent correlation between $N_{RA}$ and $E''$ for the temperature range of interest. It suggests that the RAs should be responsible for the $\beta'$ peak. To uncover the roles that VAs and RAs play in the formation of the $\beta'$ peak, we performed the atomic-pinning method [33-35] to study the effects of VAs and RAs on the $E''$, respectively. The corresponding results are shown in Fig. 5(b) after 80 atoms (only 2.0% of total atomic number, more details in [39]) from VAs or RAs are pinned. Unexpectedly, the $\beta'$ peak is completely suppressed by RA-pinning, while it's slightly suppressed by VA-pinning. Similar information is contained in $p(u')$ in Fig. 5(c), that the second peak of $p(u')$ is totally suppressed if the RAs are pinned while slightly reduced if the VAs are pinned. The results provide compelling evidence for the intrinsic link between the reversible atomistic motion and $E''$. More importantly, these also provide us the opportunity for understanding the structural origin of the $\beta'$ peak.

*3.3 Structural features of reversible atoms*

In order to describe the structural features of the RAs and get the possible correlation between the LTR behavior and the local geometry, we analyzed various local parameters (more details in [39]) of these RAs. The probability distributions of these calculated local parameters for RAs and for VAs are shown in Figs. S3, Fig. S4 and 5(d). It was found that most of (~80%) RAs are Cu with smaller atomic radius. However, of all the local structural parameters examined, only the distribution of the local configurational anisotropy (LCA, $|\Theta|$ [39]), the magnitude of the vector that points from the atom to the centroid of corresponding Voronoi cell, presents statically an obvious difference for RAs and VAs. The distribution of $|\Theta|$ of RAs shifts to larger value, which implies that the structural origin of the LTR in the MG may be due to the reversible atomic motion within the highly anisotropic atomic clusters [40].

*3.4 The influence of reversible atoms on relaxation dynamics*

Now we move one step further to clarify how the RAs identified at a given temperature could affect or be correlated with the relaxation at other temperatures, which is the key to uncover the non-monotonic evolution of $E''$. Therefore, the trajectories of the RAs identified at 300 K were also checked at 100 K and 500 K. The $p(u')$ of these atoms at the three temperatures (100 K, 300 K and 500 K) are shown in Fig. 6(a). It suggests that the RAs at 300 K may exhibit vibrational behaviors at 100 K, but diffusion-like behaviors at 500 K. To confirm that, the projected trajectories in XY plane of one RA at three temperatures are intuitively shown in Figs. 6(b)-(d). The darker color means more localized trapping and the decentralized distribution at higher temperature is due to the larger thermal fluctuation. It is noticeable that the atom is mainly trapped at two sites (site 0 and 1) at both 100 K and 300 K. As shown in Fig. 6(c) and Fig. S5 [39], the atom exhibits high frequency "jumps" between the two sites at 300 K. More importantly, the distance between the two sites is about 1.5 Å which is much smaller than the average size [~2.2 Å] of the Voronoi clusters. The geometrical evolution surrounding RA and the atomic displacements referring to site 0 are shown in the insets of Figs. 6(b) and (c). Evidently, the atomic jump between the two sites does not completely destroy the cluster or the "cage". Thus, the relaxation process may be physically understood by a double-well potential model [41]. The transition pathway between site 0 and site 1 was calculated by nudged elastic band method and shown an asymmetric behavior (Fig. S6) which can be confirmed by the asymmetric distributions of the two sites in Fig. 6(b) for 100 K. It suggests that the asymmetric double-well potential model [41-43] can also well describe the LTR process in MG. Thus, the transition frequency between two states is simply determined by $f \sim \exp(-\frac{\Delta E}{k_B T})$, where $\Delta E$ is the energy barrier and can be reduced by mechanical loading. It is worthy to mention that the inconsistent activation energy of the LTR peak between the simulation and experiment may be due to the distinct structure stability. As the temperature decreases, the $f$ decreases and the atom mostly undergoes thermal vibration by trapping within one site. However, as shown in Fig. 6(d), the sites 0 and 1 cannot be distinguished clearly and another distant site 2 appears at 500 K. It indicates that the local double-well, composed of states 0 and 1, is destroyed by thermal fluctuation, and the atomic motion leads to delocalization. The distance of site 2 is about 4.3 Å consistent with Fig. 6(a) which is comparable to the second peak of the pair distribution function g($r$) shown in Fig. 3(b) [39]. The geometrical evolution of the local cluster with atomic displacements (the insets of Fig. 6(d)) and the related trajectory (Fig. S5 [39]) indicate the breakage of the FNS, thereby causing the irrecoverable structural rearrangement. Obviously, the atom can be regarded as a FA at 500 K. Hence, we can conclude that some of the RAs at 300 K indeed originate from the VAs at 100 K, and turn into the FAs at 500 K, which results in the non-monotonic evolution of $N_{\text{RA}}$ with temperature as demonstrated in the inset of Fig. 5(a) and thereby the $\beta'$ peak. Moreover, it presents the direct evidence that the RAs served as a link connecting the thermal vibration, the low-temperature reversible relaxation and the ordinary $\beta$ relaxation caused by structural rearrangements.

*3.5 A holistic picture to understand the relaxation modes in MGs*

To understand the atomic picture of the fast relaxation peak, the spatial distribution of RAs at 300 K with displacement vectors is shown in Fig. 7(a). It presents a clear heterogeneous spatial distribution, which reflects the inherent structural heterogeneity as demonstrated in Fig. 5(d). More interestingly, the displacement vectors exhibit strong cooperative behavior at some regions as marked in Fig. 7(a). The packing cluster (Fig. 7(b)) colored by atomic displacements from one representative cooperative region confirms the "cooperative" reversible motion of the RAs at 300 K. The typical mode can be described as the oscillation of a many-atom string inside a huge cage. However, as temperature rises to 500 K, the cage is broken and the "cooperative" reversible motion evolves into the "cooperative" diffusion which supports the finding in Fig. 6(d). As proposed in the literature [19,44-48], the $\beta$ relaxation is governed by the excitations of many "cage-breaking" events through "cooperative" motions. Thus, the cooperative motions at 500 K should contribute to the $\beta$ relaxation as investigated in ref. [19]. Based on the above analysis, the intrinsic relevance between the $\beta'$ mode and the $\beta$ mode is directly evidenced by the mediation of RAs, which reveals that the $\beta'$ mode is a precursor to the $\beta$ mode.

Conceptually, our finding provides a holistic picture to understand the relaxation modes in MGs, which could be associated with the hierarchy of the PEL [Figure 7(c)]. At a very low temperature (<100 K) or very short observation time (~$10^{-12}$s), the atoms mainly vibrate around their equilibrium positions and contribute to the well-known boson peak [49]. While temperature increases (~100 K-500 K in our work) or time lengthens (~$10^{-10}$s), some atoms with relatively large LCA undergo reversible motions inside the FNSs and activate the reversible transition between two neighboring sub-basins. This contributes to the LTR peak as observed in our simulations. The relatively higher lowest-frequency mode participation ratio [50] of RAs (Fig. S4) supports the intrinsic link between vibration and fast relaxation directly. Subsequently, as these atoms escape the FNSs, the cooperative long-distance motion takes place and the β relaxation emerges [19]. When the temperature further rises to $T_\mathrm{g}$ or the observation time is increased towards the intrinsic structural relaxation time, more and more irrecoverable rearrangements are activated, resulting in the α relaxation. Finally, a critical transition from broken ergodicity to ergodicity on PEL can be observed for much higher temperature or longer time [51].

## 4. Conclusion

In summary, we reported the first pronounced low-temperature peak in the loss modulus spectrum of the model metallic glass, which bridges the gap between MD simulations and experimental studies and unravels the atomic mechanism of the related LTR peak. The extensive atomic pinning simulations reveal that the pronounced low-temperature peak originates from the reversible atomic motions inside the FNSs. By tracing the dynamic behaviors of these reversible atoms, we demonstrated that the reversible fast relaxation mode bridges the thermal vibration mode (shown as boson peak [49]) and the slow $β$ mode dominated by "cooperative" sub-diffusion [19,29]. Our work demonstrates the intrinsic hierarchy of the local relaxation modes, as triggered by the atomic vibration to the reversible and irreversible atomic movements. Moreover, the local geometrical anisotropy of the atomic clusters which accommodate the reversible atom motions is also discovered.

**Acknowledgments**

Insightful discussions with J. C. Dyre, B. Ruta, Y. J. Wang, B. Xu, Y. C. Wu, Y. C. Hu, S. Zhang are highly acknowledged. The authors acknowledge the financial support of

the NSF of China (Nos. 51571011, 51571209, 51461165101, and 51601150), the NSAF of China (No. U1530401), the MOST 973 Program (No.2015CB856800) and the Hong Kong government through the research (Nos. CityU11209317, CityU11207215). B. W. is also supported by "the Fundamental Research Funds for the Central Universities" (Grant No. G2018KY0317). We also acknowledge Beijing Computational Science Research Center (CSRC) for the computational support.

**Figures and captions**

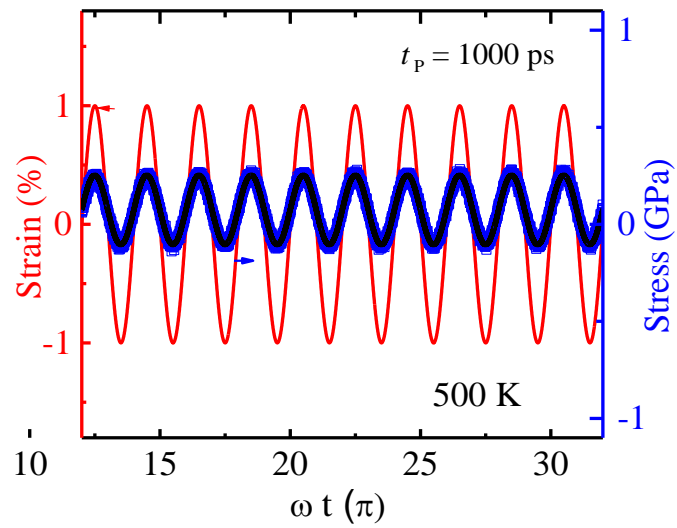

**Fig. 1.** The MD-DMS data at T=500 K with $t_p = 1000$ ps. (red, left axis) The loaded sinusoidal strain. (blue, right axis) The resultant stress fitting by a sinusoidal function (black, right axis).

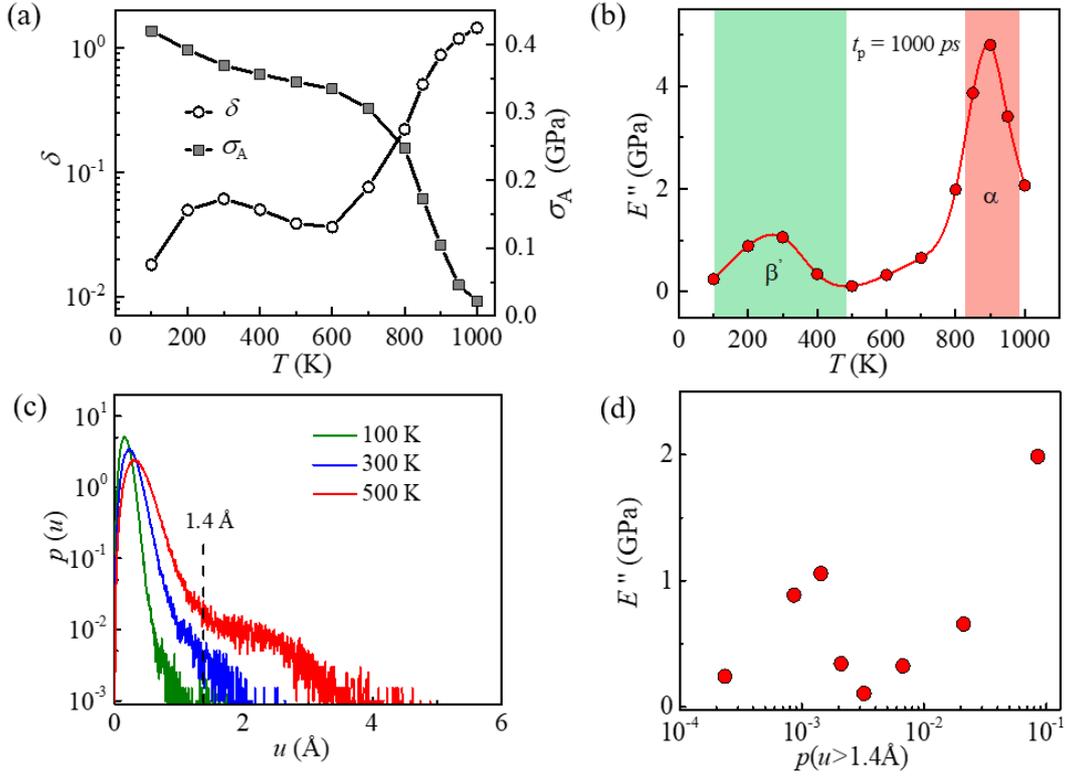

**Fig. 2** (a) The fitted $\sigma_A$ and $\delta$. (b) The calculated $E''(T)$. (c) Probability density function $p(u)$ using the conventional method $u(\Delta t) = ||\vec{r}(t_0 + \Delta t) - \vec{r}(t_0)||$, where $\Delta t = T_p = 1000$ ps. $p(u)$ is defined by the formula $p(u) = [P(u + \Delta u) - P(u)]/\Delta u$, where $P(u)$ represents the cumulative distribution quantifying the probability of finding X⩽$u$ and $\Delta u$ was set to 0.01 Å. (d) The loss modulus dependence of the fraction of atoms with u>1.4 Å.

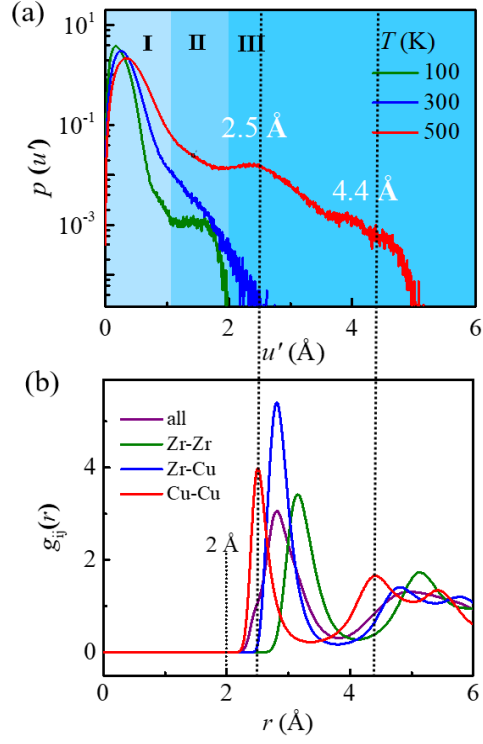

**Fig. 3** (a) Probability density function $p(u')$ at 100 K, 300 K and 500 K with displacement $u'$ calculated by $u'(t) = \left\| \vec{r}(t) - \vec{r}(6T_p) - \int_{6T_p}^{t} \dot{\varepsilon} y(t)\hat{x} dt \right\|$ over a time range, $t \in [6T_p, 16T_p]$. (b) Pair distribution functions $g_{ij}(r)$ for different atomic pairs at 500 K. (c) $p(u')$ of VAs at 100 K.

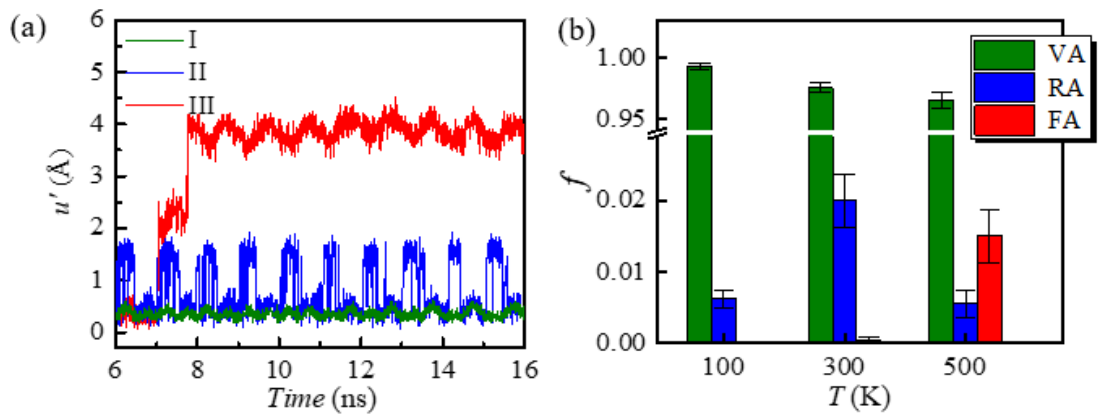

**Fig. 4** (a) Time evolution of $u'$ for the three typical atoms. (b) The fraction of each type of atoms in (a) during cycling.

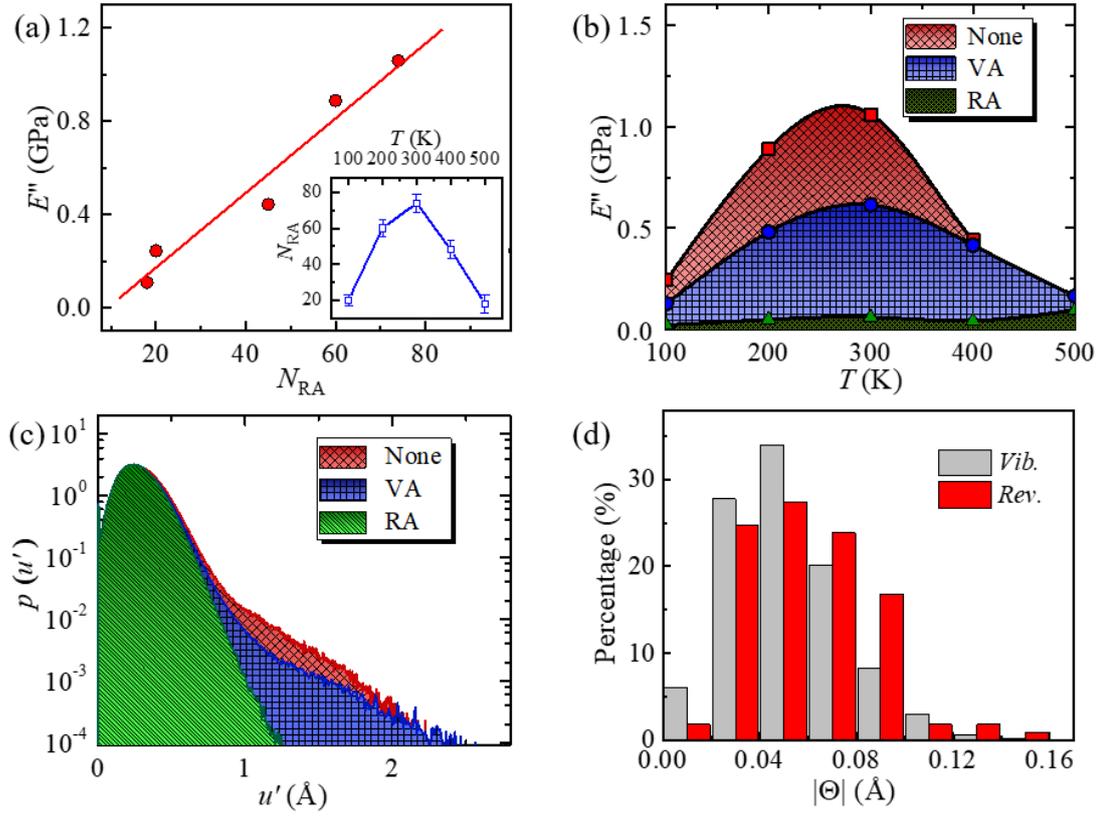

**Fig. 5.** (a) The number of reversible atoms ($N_{RA}$) dependent loss modulus ($E''$) at low temperatures. Inset: the $N_{RA}$ as a function of temperature. (b) The measured $E''$ under the conditions that no atoms, reversible atoms (RAs) and vibrational atoms (VAs) are pinned, respectively. The atoms pinned only move affinely with the simulation box. (c) The $p(u')$ for various pinning situations in (b). (d) The distributions of the local configurational anisotropy (LCA, $|\Theta|$ [39]), where LCA represents the magnitude of the vector pointing from the atom to the centroid of its Voronoi cell, for VAs and for RAs.

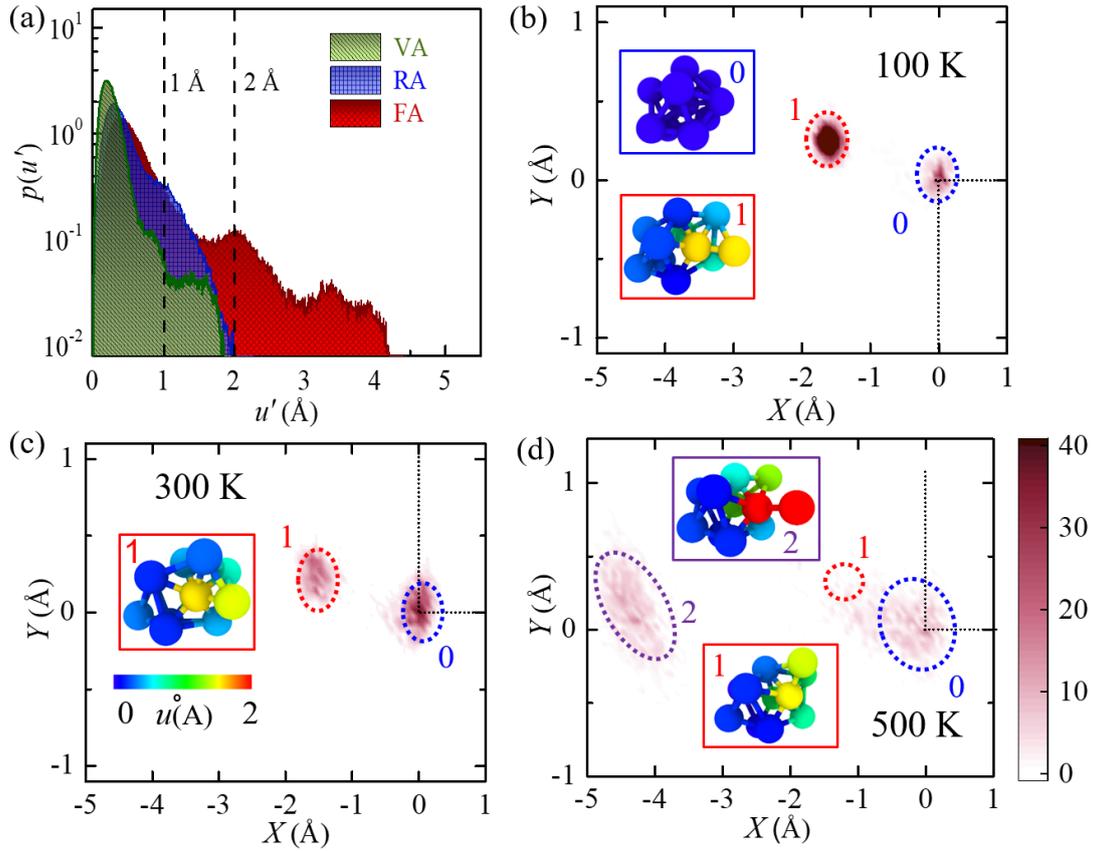

**Fig. 6.** (a) The $p(u')$ of selected RAs at the three temperatures. The projected trajectories in XY plane of one RA at (b) 100 K, (c) 300 K and (d) 500 K. The initial positions are shifted to (0,0) and the gradient purple color bar presents spatial probability distribution. The dotted ellipses exhibit the most probable or representative places where the atom appears. Inset: the related local clusters with atoms colored by displacement. In (b) shows the initial set of displacements labeled "0", which is omitted in (c) and (d).

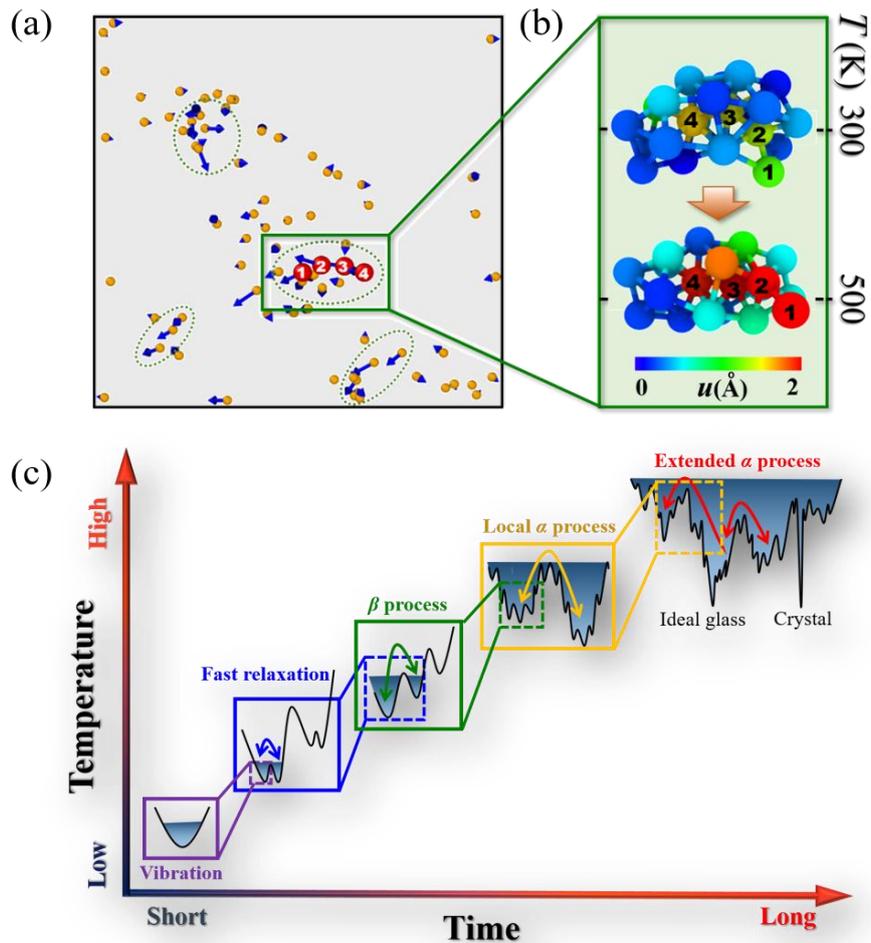

**Fig. 7.** (a) Two-dimensional schematic of RAs at 300 K. The blue arrows exhibit atomic displacement vectors. The pronounced "cooperative" motions are labelled by dotted ellipses. (b) Local clusters surrounding the labelled atoms in (a) at 300 K and 500 K, colored by displacement. (c) The holistic picture of the relaxation modes in MGs based on the hierarchy of the PEL.

# Supplemental Material for

# Revealing the low-temperature fast relaxation peak in a model metallic glass

B. Wang[a,b,$], L. J. Wang[c,d,$], B. S. Shang[c], X. Q. Gao[e], Y. Yang[f,*], H. Y. Bai[a,g,h], M. X. Pan[a,g,h *], W. H. Wang[a,g,h], P. F. Guan[c,*]

[a]Institute of Physics, Chinese Academy of Sciences, Beijing 100190, China

[b]School of Physical Science and Technology, Northwestern Polytechnical University, Xi'an 710072, China

[c]Beijing Computational Science Research Center, Beijing 100193, China

[d]School of Physics and Materials Science, Anhui University, Hefei 230601, China.

[e]Northwest Institute for Nonferrous Metal Research, Xi'an 710016, China

[f]Department of Mechanical Engineering, City University of Hong Kong, Tat Chee Avenue, Kowloon Tong, Kowloon, Hong Kong, China

[g]School of Physical Sciences, University of Chinese Academy of Sciences, China

[h]Songshan Lake Materials Laboratory, Dongguan, Guangdong 523808, China

## 1. The distribution density functions $p(u')$ at 100 K for vibrational atoms

The non-affine displacement, $u'(t) = \left\|\vec{r}(t) - \vec{r}(6T_p) - \int_{6T_p}^{t} \dot\varepsilon y(t)\hat{x}dt\right\|$ with $\dot\varepsilon = 2\pi\varepsilon_A/T_p \cos(2\pi t/T_p)$, over the time range, $t \in [6T_p, 16T_p]$ was calculated at 100 K for vibrational atoms. Here we start the analysis after 6 cycles to eliminate the artificial unstable local atomic packing and to ensure the system has reached the steady state under the sinusoidal loading. The $p(u')$ shall be calculated in the same way as $p(u)$, shown in Fig. S1.

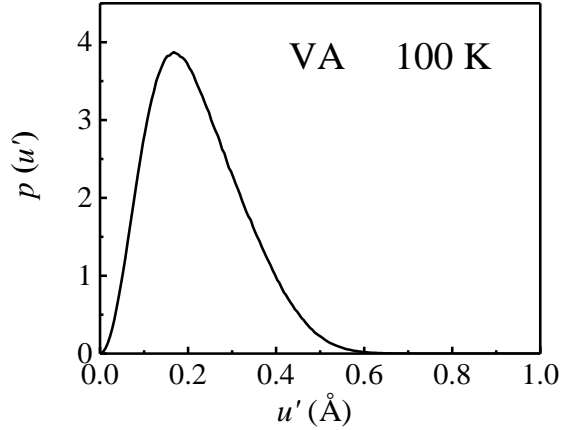

**Fig. S1** (c) $p(u')$ of VAs at 100 K.

## 2. Selection of RAs

We performed 50 independent MD-DMS simulations based on the isoconfigurational ensemble to analyze the motion for each atom. During the whole loading process, 500 cyclic periods were considered. In one cyclic loading period, if the reversible motion can occur, it is counted as once. Therefore, the maximum of times the reversible motion of one atom occurs in 500 cyclic loadings can only be 500. Here, we define that if the times of the reversible motion of one atom exceed 250 during 500 MD-DMS cyclic loading processes, then the atom can be considered as a RA. In other words, if the atom has the probability of more than 50% to behave as reversible motion, the atom can be considered as a RA.

## 3. The features of local parameters for RAs and VAs

In order to characterize the structural features of RAs, we calculated the free volume (V), bond orientational order ($Q_6$), potential energy (E), atomic level stress (S) and local configurational anisotropy $|\Theta|$ [1-5]. The followings are the calculation methods of these parameters and the comparison of these parameters between VAs and RAs.

**Free volume V.** Voronoi tessellation method often was used to depict the random dense structure of metallic glasses, and the Voronoi volume of each atom was employed to get the free volume V [4]. In the method, the Voronoi polyhedron is composed of perpendicular bisection planes between center atom and its neighbor atoms. In a simple two dimensional (2D) system, the sketch of Voronoi polygon is shown in Fig. S2(a).

**Bond orientational order (BOO).** The bond orientational order of atom $i$ can be represented by $q_l^i = [\frac{4\pi}{(2l+1)}\sum_{m=-l}^{l}|q_{lm}^i|^2]^{1/2}$, and $q_{lm}^i = \frac{1}{n}\sum_{j=1}^{n} Y_{lm}(\vec{r}_{ij})$, where $Y_{lm}(\vec{r}_{ij})$ is a spherical harmonic function of degree $l$ and order $m$, $n$ is the number of the nearest neighbors of particle $i$. Then its rotational invariant combination can be extracted according to $Q_l^i = [\frac{4\pi}{(2l+1)}\sum_{m=-l}^{l}|\bar{q}_{lm}^i|^2]^{1/2}$, where $\bar{q}_{lm}^i = \frac{1}{N}\sum_{j=1}^{N} q_{lm}^j$, and the sum runs over all neighbors of particle $i$ plus particle $i$ itself [6]. Here we used $Q_l^i$ to calculate $Q_6$.

**Potential energy (E).** The potential energy (E) of each atom can be extracted by opening LAMMPS directly.

**Atomic level stress (S).** The von Mises stress of atom $i$, $\sigma_{VM}^i = \sqrt{\frac{1}{2}[(\sigma_{11}^i - \sigma_{22}^i)^2 + (\sigma_{22}^i - \sigma_{33}^i)^2 + (\sigma_{11}^i - \sigma_{33}^i)^2] + 3[(\sigma_{12}^i)^2 + (\sigma_{13}^i)^2 + (\sigma_{23}^i)^2]}$, from virial stress tensors was calculated to represent S [7].

**Local configurational anisotropy |Θ|.** We know that the position of a particle within its Voronoi cell is an indicator of local variation in the packing. When configuration is disordered, the center of the particle deviates from the centroid of corresponding Voronoi cell, and the magnitude of Voronoi cell anisotropy vector that points from the center of the particle to the centroid can represent the extent of structural anisotropy, denoted as the local configurational anisotropy (LCA) [5]. Here, the centroid of Voronoi cell is the mean position of all the points of Voronoi cell in all of the coordinate directions. For example, for simple 2D triangle, its centroid is the intersection of the three medians of the triangle (each median connecting a vertex with the midpoint of the opposite side shown in Fig. S2(b) by yellow dotted line). And for 2D Voronoi polyhedron S in Fig. S2(b), the centroid can be computed by dividing polyhedron into a finite number of simple figures $S_1, S_2, \cdots, S_N$, computing the centroid $S_i$ and area $A_i$ of each part, and then computing $S_x = \frac{\sum S_{ix} A_i}{\sum A_i}$, $S_y = \frac{\sum S_{iy} A_i}{\sum A_i}$. Based on above definition and calculation, the red solid circle in Fig. S2(b) is the centroid of the Voronoi cell which resides one particle-the blue solid circle, and Voronoi cell anisotropy vector $\vec{\Theta}$ is defined as the vector that points from the center of the particle to the centroid of its Voronoi cell (Fig. S2(c)). In 3D polyhedrons, the computing method of centroid is the

same as in 2D above.

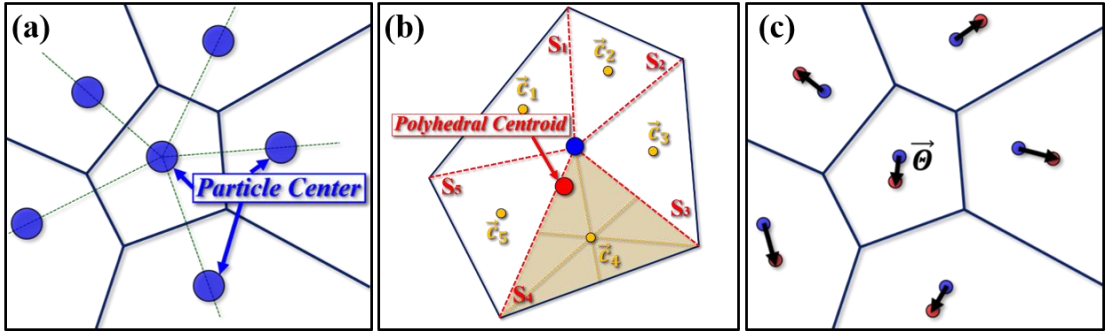

**Fig. S2** The sketch map of Voronoi polyhedron and the Voronoi cell anisotropy vector. (a) The schematic diagram of Voronoi polyhedron. The center particle has five first nearest neighbor atoms. The green dotted lines mean the perpendicular bisectors between the centers of adjacent particles, and blue polygons correspond to the Voronoi cell. (b) The centroid of Voronoi cell. For a simple triangle, its centroid ($c_4$) is the intersection of the three medians of the triangle shown as yellow dotted lines (each median connecting a vertex with the midpoint of the opposite side). For a complex Voronoi cell S, the centroid of it can be computed by dividing it into a finite number of simple figures $S_1, S_2, \cdots, S_5$, computing the centroid $S_i$ and area $A_i$ of each part, and then computing $S_x = \frac{\sum S_{ix} A_i}{\sum A_i}$, $S_y = \frac{\sum S_{iy} A_i}{\sum A_i}$. The red circle is the centroid of the Voronoi cell. (c) the Voronoi cell anisotropy vector $\vec{\Theta}$ is defined as the vector that points from the center of the particle to the centroid of its Voronoi cell.

The probability distributions of these calculated local parameters for RAs and for VAs are shown in Figs. S3 and 5(d). The volume and energy distributions are bimodal, which indeed corresponds to Cu and Zr atoms. The reversible atoms mostly are Cu atoms, which were confirmed by the selected reversible atoms' type. For structural parameters, only the distribution of the local configurational anisotropy (LCA, |Θ|) for RAs presents statically an obvious tendency to higher values if compared to that for VAs, consistent with the change of potential energy and atomic level stress.

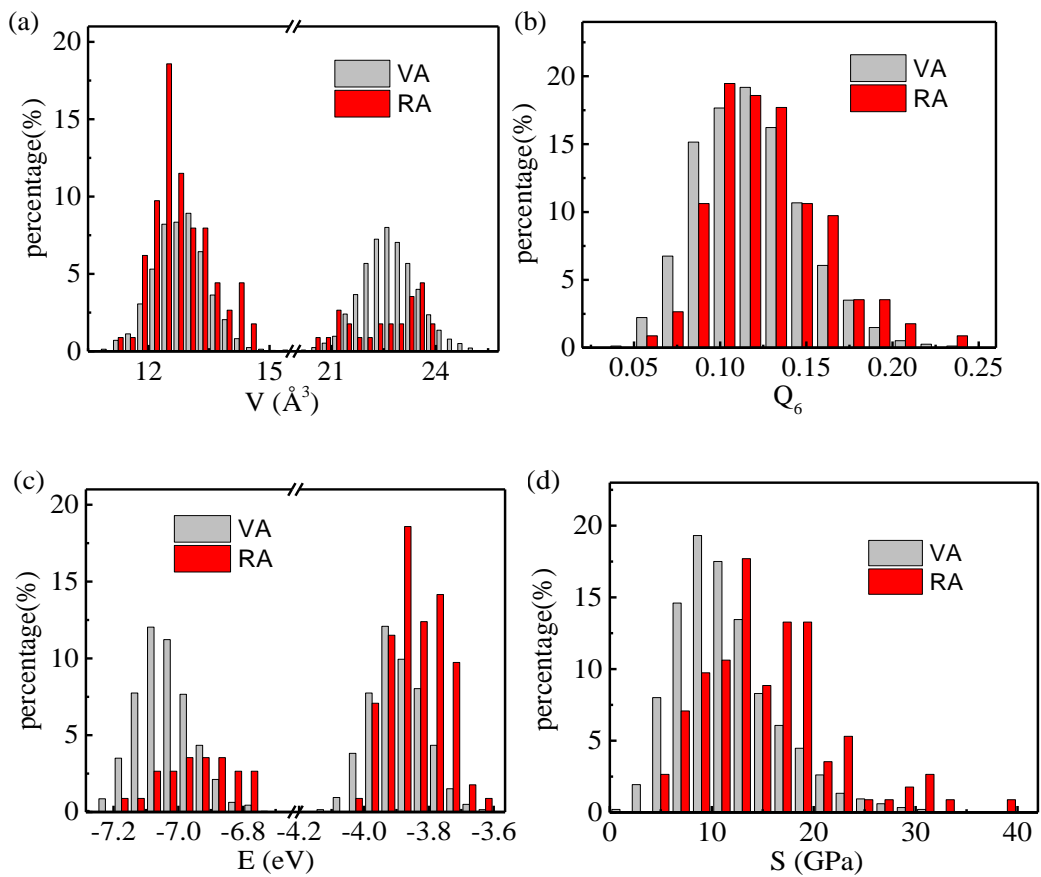

**Fig. S3** The histogram of (a) free volume (V), (b) bond orientational order $Q_6$, (c) potential energy (E) and atomic level stress (S) for VAs and for RAs.

**Participation ratio of lowest-frequency modes**. The normal modes were calculated by diagonalizing the dynamic matrix of the MG inherent structure. The participation ratio of particle $i$ in the eigenmode $\vec{e}_\omega$ with frequency ω was calculated by $p_i = |\vec{e}_\omega^i|^2$ with $\vec{e}_\omega^i$ the corresponding polarization vector of particle $i$ in $\vec{e}_\omega$. The average participation ratio of the $i$th atom denoted as $P_i$ was averaged over the 1% lowest-frequency normal modes and a larger value of $P_i$ is usually indicative of larger vibrational amplitude of atom $i$. The reversible atoms tend to correlate with soft spots with large participation ratios, as shown in Fig. S4. It provides the strong evidence for the intrinsic link between the atomic vibration and revisible motion in our work.

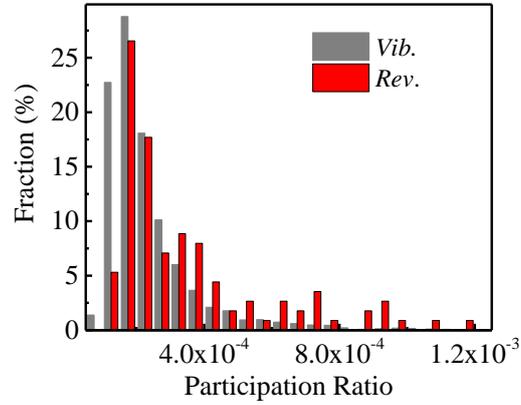

**Fig. S4.** The participation ration of each atom is averaged over the 1% lowest-frequency modes (soft modes), which measures the vibrational amplitude of that atom. The figure shows the histogram of participation ratio of RAs and VAs.

## 4. Time evolution of atomic displacements for the RA in Fig. 3

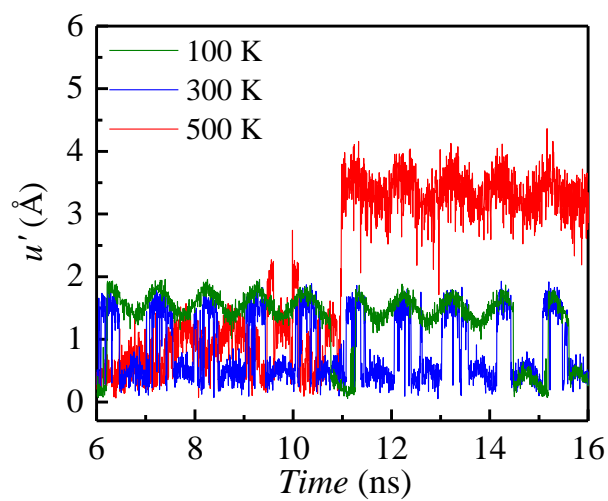

**Fig. S5** Time evolution of atomic displacements for the RA in Fig. 3 at the three temperatures (100 K, 300 K and 500 K).

## 5. Activation Energy of revisable motions

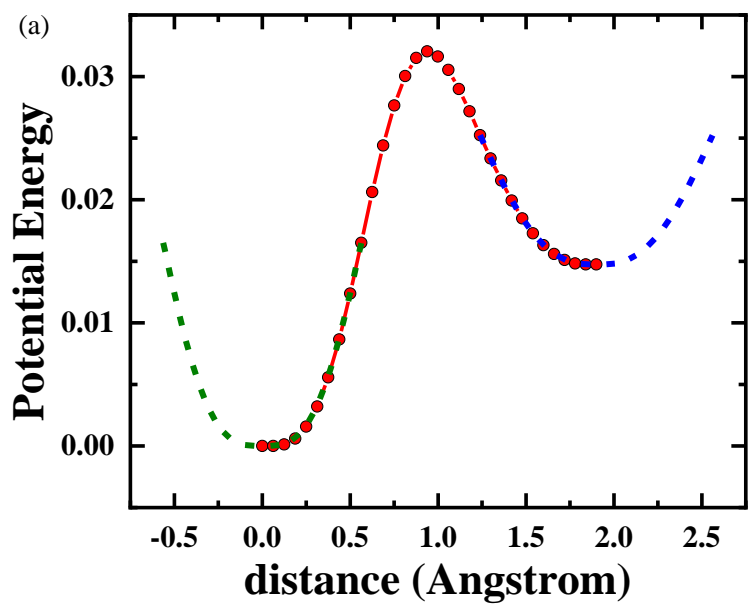

(a)

(b)

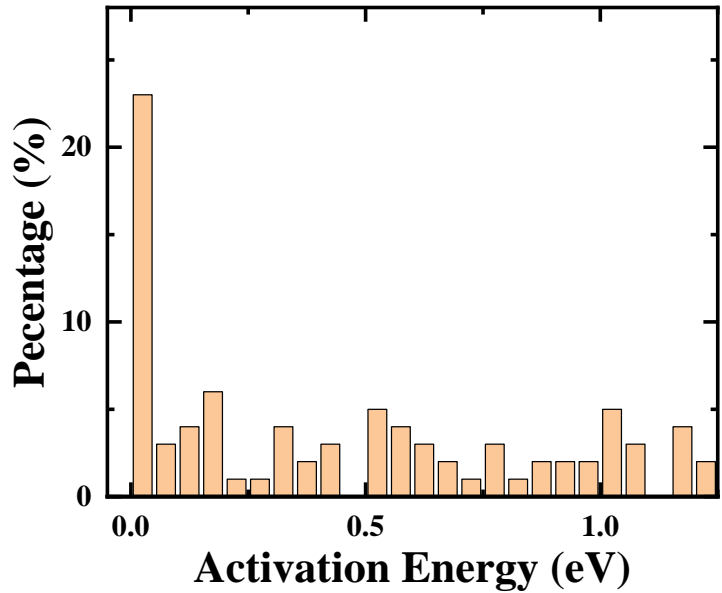

**Fig. S6**. (a) One example activation event of a RA. (b) Distribution of activation energies of RAs.